\documentclass[a4paper,11pt]{article}
\usepackage[english]{babel}

\usepackage[english]{babel}
\usepackage[utf8]{inputenc} 
\usepackage{graphicx} 
\usepackage{fancyhdr}
\usepackage{amsmath}
\usepackage{physics}
\usepackage{slashed}
\usepackage{dingbat}

 \usepackage{cite}
\usepackage{mathptmx}
\usepackage{stackengine}
\usepackage{mattens}
\usepackage{eucal}
\usepackage[usenames, dvipsnames]{color}

\begin{document}

\begin{center} 
\vspace*{1.5cm} 
 
{\Large{\textbf {{Exploring Critical Overdensity Thresholds in Inflationary Models of Primordial Black Holes Formation}}}}\\ 
 \vspace*{10mm} 
 
 {\bf  Ioanna~D. Stamou} $^1$ \\ 
  
\vspace{.7cm} 

 $^1${\it Service de Physique Th\'  eorique, C.P. 225, Universit\'e Libre de Bruxelles,\\
Boulevard du Triomphe, B-1050 Brussels, Belgium}

\end{center} 
\vspace{2.cm}

\begin{abstract}
In this paper we study the production of Primordial Black Holes (PBHs) from inflation in order to explain the Dark Mater (DM) in the Universe. 
The evaluation of the fractional PBHs abundance to DM is sensitive to the value of the  threshold $\mathrm{\delta_c}$ and the exact value of  $\mathrm{\delta_c}$ is sensitive to the specific shape of the cosmological fluctuations. Different mechanisms  producing PBHs lead to different thresholds and hence to different fractional abundances  of PBHs. 
In this study, we  examine various classes of inflationary models proposed in the existing literature to elucidate the formation of PBHs and we evaluate numerically the associated threshold values.
 Having evaluated the thresholds we compute the abundances of PBHs to DM using the Press Schecter approach and  the Peak Theory. Given the influence of different power spectra on the thresholds, we investigate whether these inflationary models can successfully account for a significant fraction of DM. Moreover, we provide suggested values for the critical threshold. 
By examining the interplay between inflationary models, threshold values, and PBH abundances, our study aims to shed light on the viability of PBHs as a  candidate for DM and contributes to the ongoing discussion regarding the nature of DM in the Universe.

\end{abstract}

\section{Introduction}

Dark Matter (DM) is considered as one of the biggest problem in Cosmology. Recent  observations, such as the detection of Gravitational Waves emitted by a binary black hole merger \cite{Abbott:2016blz,Abbott:2017vtc,Abbott:2017gyy,Abbott:2017oio,Abbott:2016nmj}, have reignited interest in the possibility that Primordial Black Holes (PBHs) could constitute a significant fraction of DM. The idea of PBHs was proposed in the 1970s by Hawking and Carr \cite{Hawking:1971ei,Carr1974}. The renewed detection of Gravitational Waves has revitalized the exploration of the connection between DM and PBHs.

In particular, there are numerous  theoretical studies on the formation of PBHs from inflationary models, such as Refs. \cite{Braglia:2020eai,Clesse:2015wea,Spanos:2021hpk,Ballesteros:2017fsr,Gao:2018pvq,Cicoli:2018asa,Dalianis:2018frf,Garcia-Bellido:2017mdw,Ezquiaga:2017fvi,Nanopoulos:2020nnh,Stamou:2021qdk,Hertzberg:2017dkh,Ballesteros:2019hus,Mahbub:2019uhl,Aldabergenov:2020bpt,Stamou:2021vms,Meng:2022low,Braglia:2020fms,Geller:2022nkr,Braglia:2020taf,Kallosh:2022vha,Mavromatos:2022yql,Boutivas:2022qtl,Braglia:2022phb,Tada:2023pue,Ashoorioon:2019xqc,Ashoorioon:2020hln,Ashoorioon:2022raz,Belotsky:2018wph,Khlopov:2008qy,Spanos:2022euu,Kawai:2022emp,Kawai:2021edk,Ferrante:2022mui,Poisson:2023tja}.  According to these studies an enhancement in the power spectrum at small scales can lead to PBHs formation, which could  explain a significant fraction of DM (or even the whole DM)  in the Universe.  Many of these models are based on single field inflation   with  a near inflection point in the  scalar potential, such as those of Refs.~\cite{Ballesteros:2017fsr,Gao:2018pvq,Cicoli:2018asa,Dalianis:2018frf,Garcia-Bellido:2017mdw,Ezquiaga:2017fvi,Nanopoulos:2020nnh,Stamou:2021qdk,Hertzberg:2017dkh,Ballesteros:2019hus,Mahbub:2019uhl,Aldabergenov:2020bpt,Stamou:2021vms}. The drawback of these models is that a lot of fine-tuning in the underlying parameters is required. 
Other models with two-field inflation have been proposed~\cite{Braglia:2020eai,Meng:2022low,Braglia:2020fms,Geller:2022nkr,Braglia:2020taf,Kallosh:2022vha,Mavromatos:2022yql,Boutivas:2022qtl}. For instance, hybrid models have  been intensely  studied in the  literature~\cite{Clesse:2015wea,Spanos:2021hpk,Braglia:2022phb,Tada:2023pue}. 
The formation of PBHs in the majority of these models takes place in the radiation dominated epoch. 
In our study we only consider this class of models. 
 

The formation of PBHs  occurs when a cosmological perturbation collapses to a black hole if its amplitude  $\mathrm{\delta}$ exceeds a certain threshold value $\mathrm{\delta_c}$. Early analytical estimates of $ \mathrm{\delta_c}$ were based on a simplified Jeans length approximation, which gives $\mathrm{\delta_c} \sim w$, where $\mathrm{w}$ is the equation of state \cite{1975ApJ...201....1C}. 
 More recent studies have refined this value by incorporating the theory of General Relativity, obtaining  $ \mathrm{\delta_c \sim 0.4	} $  in the radiation dominated era \cite{Harada:2013epa}. However, this analytical computation provides only a lower bound because it does not account for non-linear effects. Full numerical relativistic simulations are required to fully capture these effects, which recent studies have shown to be dependent on the initial curvature profile, with $\mathrm{0.4 \leq	 \delta_c \leq	 2/3}$ \cite{Musco:2020jjb,Germani:2018jgr,Musco:2018rwt,Escriva:2019phb,Escriva:2022duf,Musco:2021sva,Yoo:2020lmg,Kehagias:2019eil}. 
Significant progress has been made in understanding the mechanism of PBH formation through detailed spherically symmetric numerical simulations that incorporate a non-linear approach\cite{Shibata:1999zs,Escriva:2020tak,Harada:2015yda,Escriva:2019nsa,Musco:2012au,Musco:2008hv,Musco:2004ak,IHawke2002,Niemeyer:1997mt,Bloomfield:2015ila,Escriva:2021aeh}.

It was previously remarked that 
the threshold for PBHs formation depends on the specific mechanism of inflation and the properties of the collapsing object, such as its mass, size, and initial density profile \cite{Harada:2015yda,Shibata:1999zs,Escriva:2020tak,Musco:2004ak,Musco:2018rwt,Carr:2009jm,Carr:2020gox}. 
In addition to that, the abundances of DM from PBHs are extremely sensitive to the threshold and  the exact value of the peak.  
In other words, the shape of the power spectrum leads to a different threshold and hence to a different abundance of PBHs to DM~\cite{Germani:2018jgr}. 
\textcolor{black}{ In the aforementioned inflationary models  an acceptable value of the threshold is used and the fact that the critical threshold is depended on the shape of the power spectrum was neglected. As the evaluation of the fractional abundance of PBHs is crucial depended on the threshold and this threshold depends on the power spectrum, we believe that it should be tested that these models can indeed explain a significant fraction of  DM.  Therefore, as the exact value of the threshold has an important role in the calculation, it can lead to  either ruling out or not  inflationary models for producing PBHs as DM. }


In this study we evaluate numerically the threshold for some classes of inflationary models presented in the literature. Specifically, we study the thresholds for the case of two inflation model with a non-canonical kinetic term \cite{Braglia:2020eai}, the case of hybrid model~\cite{Clesse:2015wea,Spanos:2021hpk} and the case of a single field with an inflection point\cite{Stamou:2021qdk}. Having these values for the threshold we evaluate the abundance of PBHs to DM. Instead of having an acceptable value for the threshold, we evaluate numerically these thresholds for each case. 
\textcolor{black}{The advantage of these three models is that they allow the calculation of the evaluation of the critical threshold for just one parameter. One parameter dependence is needed in order to avoid a complicated numerical evaluation of the threshold from a power spectrum of a given model. Other models provided in the literature with same shapes of the power spectrum may not lead to one-to-one comparison of the shape and the critical threshold, as many parameters can change the results. However, the results of the critical value can be applied to the other models with similar shape of power spectrum. In this study, we can conclude if these models can predict a significant fraction of DM and we can propose a value for the threshold for those models. }

The layout of this paper is as follows: in Section 2 we   present the basic aspects  for the threshold $\mathrm{\delta_c}$. In  Section 3 we show the evaluation of $\mathrm{\delta_c}$ for a given power spectrum. In Section 4 we  present the calculation of the abundances of PBHs. In Section 5 we present the application of the previous analysis to inflationary models and especially to two field models, hybrid models and models with an inflection point. Finally, we draw our conclusions in Section 6.

\section{The threshold $\mathrm{\delta_{c}}$}
In this section we introduce the threshold of PBHs. Generally, the threshold for PBHs  is determined by the density fluctuations in the early Universe, which can collapse under their own gravity to form black holes if they exceed a certain threshold value.

The PBHs are formed from cosmological fluctuations after  re-entering the horizon. Under the assumption of spherical symmetry, the spacetime metric on the superhorizon scales can be given as
\begin{equation}
ds^2=-dt^2+a^2(t)\left[ \frac{dr^2}{1- K(r)r^2} +r^2 d\Omega^2\right]=-dt^2+a(t)^2\exp(2\zeta(\hat{r}))[d\hat{r}^2+\hat{r}^2d\Omega^2]
\end{equation}
where $a\mathrm{(t})$ is the scale factor and $\mathrm{K}$ and $\mathrm{\zeta(r)} $ are the conserved comoving curvature perturbations defined at the superhorizon scales.
Combining these expressions we have:
\begin{equation}
K(r)r^2=-\hat{r}\zeta'(\hat r) (2+\hat{r} \zeta'(\hat{r})).
\end{equation}
and the coordinates $\mathrm{r}$ and $\mathrm{\hat r}$
are connected as follows:
\begin{equation}
\begin{split}
r&= \hat{r}\exp\left(\zeta(\hat{r} )\right)\\
&\frac{dr}{\sqrt{1-K(r)r^2}}=\exp\left(\zeta(\hat{r} )\right)d \hat{r}.
\end{split}
\end{equation}
As shown in \cite{Harada:2015yda,Shibata:1999zs,Escriva:2020tak,Musco:2004ak,Musco:2018rwt}  the shape of cosmological perturbations  is related to the $\mathrm{K}$ and $\mathrm{\zeta(r)}$ and so they are related to the power spectrum $\mathrm{P_R}$. In other words, different scalar power spectrum profiles can result in varying thresholds.

The energy density profile is defined from:
\begin{equation}
\frac{\delta \rho}{\rho_b} \equiv	\frac{\rho(r,t)-\rho_b(t)}{\rho_b(t)}= f(w) \left( \frac{1}{aH}\right)^2 \left( K(r) +\frac{r}{3}K'(r)\right)
\end{equation}
where  $\mathrm{H}$ is the Hubble parameter,  ${\rho_b}$ is the mean background energy density and $\mathrm{f(w)}$ is:
\begin{equation}
\mathrm{f(w)= 3(1+w)/(5+3w)}
\label{eq:fw}
\end{equation}
with $\mathrm{w}$ is the equation of state $\mathrm{w=p/\rho}$. 

The criterion to define the PBHs formation can be given at the peak of the compactification function, which is defined as follows:
\begin{equation}
C\equiv2\frac{\delta M}{R(r,t)}
\end{equation}
where $\mathrm{R}$ is the areal radius and $\mathrm{\delta M= M(r,t)- M_b (r,t)}$. $\mathrm{M}$ is the Misner-Sharp mass and $\mathrm{M_b(r,t)=4\pi \rho_bR^3/3}$.  
If we define the expansion parameter $\varepsilon$  as follows
\begin{equation}
\varepsilon=\frac{1}{a(t)r_mH(t)}=\frac{1}{a(t) \hat{r}_m\exp(\zeta) H(t)}
\end{equation}
one can express the compactification function $\mathrm{C}$ at the leading order $\mathrm{O(\varepsilon^3) }$  of the gradient expansion of  Misner-Sharp equations as follows \cite{Harada:2015yda,Polnarev:2006aa,Polnarev:2012bi,Escriva:2022duf}:
\begin{equation}
C \simeq	f(w) K(r)r^2=f(w)\left( 1- [1+\hat r \zeta'(\hat r)]^2\right).
\end{equation}
The expression of energy density profile is valid if $\mathrm{\varepsilon \ll 1}$. 
Finally, the threshold $\mathrm{\delta_c}$ is equivalent to the peak value of the compactification function, $\mathrm{\delta_c}=\mathrm{C({\hat r}_m)}$. Fluctuations with  with amplitude $\mathrm{\delta_m}$ bigger than the threshold,  $\mathrm{\delta_m> \mathrm{\delta_c}}$, can collapse to form PBHs, otherwise  fluctuations  with $\mathrm{\delta_m< \mathrm{\delta_c}}$ cannot lead to PBHs formation.

Having the assumption of spherically symetric perturbation we define the dimensionless shape parameter $\mathrm{q}$\cite{Escriva:2019phb,Musco:2018rwt}  as: 
\begin{equation}
q=-\frac{C''( r_m){{ r}_m}^2}{4C( r_m)}.
\label{eq:q_shape_parameter}
\end{equation}
and in terms of $\mathrm{\hat r}$ it takes the form:
\begin{equation}
q=-\frac{C''({\hat r}_m)\hat{r}_m^2}{4C({\hat r}_m)\left( 1-\frac{C(\hat r_m)}{f(w)}\right)}.
\end{equation}
This parameter is characterised from the width of the peak of the compactification function. 
We remark here that different profiles with same parameter $\mathrm{q}$ have the approximated same threshold $\mathrm{\delta_c}$ in the case of radiation epoch \cite{Escriva:2019phb}.

\section{The estimation of the threshold}
\label{The evaluation of the threshold}

There is a  number of works where an analytical expression for the threshold $\delta_c$ was studied~\cite{Carr1974,Escriva:2019phb,Musco:2020jjb}. In our study we use this of Ref.\cite{Escriva:2019phb}.  
In this section, we summarize the evaluation of the threshold for a given scalar power spectrum $\mathrm{P_R}$. As we remarked previously, the shape of the power spectrum leads to a different threshold. This analysis is presented in Refs.\cite{Musco:2020jjb,Escriva:2022duf,Escriva:2019phb,Musco:2018rwt,Germani:2018jgr}. 

In general, the power spectrum is given by 
\begin{equation}
P_R(k)=\frac{k^3}{2\pi^2}|\mathcal{R_k}|^2
\end{equation}
with $\mathrm{\mathcal{R_k}}$ the curvature perturbation and $\mathrm{k}$ is the comoving wavenumber. 
 A concise analysis of the power spectrum evaluation will be presented later.
In this section we follow the steps of Ref. \cite{Escriva:2022duf}  in order to obtain the threshold $\mathrm{\delta_{c}}$ for a given power spectrum. An equivalent analysis is presented in Ref. \cite{Musco:2020jjb}.

In order to obtain the threshold, one should consider  the two point correlation function as:
\begin{equation}
g( {\hat r})=\frac{1}{{\sigma}_0^2}\int^{\infty}_{-\infty} \frac{dk}{k}\frac{sin(k \hat r)}{k{\hat r}}P_R(k).
\label{eq:g_two_point}
\end{equation}
with 
\begin{equation}
\sigma_0^2=\int^{\infty}_{-\infty}dk \frac{P_R(k)}{k}.
\end{equation}
The Eq.~(\ref{eq:g_two_point}) connects the scalar power spectrum $\mathrm{P_R}$  with the two point correlation function.

As a first step  in the calculation of the $\mathrm{\delta_{c}}$,  we should locate the maximum value of the compactification function $\mathrm{\hat r_m}$. The value of $\mathrm{\hat r_m}$ can be obtained by the root of the following  equation:
\begin{equation}
\zeta'(\hat r_m)+\hat r_m\zeta''(\hat r_m)=0
\end{equation}
where $\mathrm{\mu}$ is the amplitude of curvature fluctuation and it is given as follows:
\begin{equation}
\mu= \zeta(\hat  r)/g(\hat  r)
\end{equation}
or
\begin{equation}
\mu=\frac{\pm\sqrt{1-C(\hat  r_m)/f(w)}-1}{g'(\hat  r_m)\hat  r_m}.
\end{equation}
The critical amplitude, $\mathrm{\mu_c}$, is obtained at $\mathrm{C(r_m)=\delta_c}$.

In order to obtain the threshold we define the function
\begin{equation}
G(\hat{r}_m)=\frac{g'(\hat{r}_m)-{\hat{r}_m}^2 g'''(\hat{r}_m)/2}{g'(\hat{r}_m)}.
\end{equation}
The threshold $\delta_{c}$ can be given from the following expressions:
\begin{equation}
q=G(r_m)\frac{1}{\sqrt{1-\delta_c(q)/f(w)}} \frac{1}{(1+ \sqrt{1-\delta_c(q)/f(w)})}
\label{eq:shapeq}
\end{equation}
where $\mathrm{\delta_c}$ is given:
\begin{equation}
\delta_{c}=\frac{4}{15}e^{-1/q}\frac{q^{1-5/2q}}{\Gamma(\frac{5}{2q})-\Gamma(\frac{5}{2q},\frac{1}{q})}.
\label{eq:dc}
\end{equation}
The dimensionless parameter $\mathrm{q}$ given in Eq.~(\ref{eq:shapeq}) is the shape parameter and it is defined as the shape around the peak of  the  compactification function given previously  in Eq.(\ref{eq:q_shape_parameter}).
The Eq.~(\ref{eq:dc}) is an analytical expression to calculate the threshold $\mathrm{\delta_c}$  as a function of the shape parameter $\mathrm{q}$  in radiation era\cite{Escriva:2019phb}.  Finally, $\mathrm{\Gamma}$ is the incomplete  gamma function.   

\section{The PBHs production}

Until now, we have introduced the evaluation of the thresholds $\mathrm{\delta_c}$. As we study the possibility of PBHs to explain the DM, we introduce
 the evaluation of the fraction of PBHs to DM,  $\mathrm{f_{PBH}}$. In this section we summarize the calculation of  $\mathrm{f_{PBH}}$. 

The present abundance of PBHs is given by the integral:
\begin{equation}
f_{PBH} = \int d\ln M \frac{\Omega_ {PBH}}{\Omega_ {DM}}
\label{eq:abundance}
\end{equation}
\noindent
and the fractional abundance of PBHs to DM is:
\begin{equation}
\label{eq:abundance1}
\frac{\Omega_ {PBH}}{\Omega_ {DM}}= \frac{\beta(M_{PBH}(k))}{8 \times 10^{-16}} \left(\frac{\gamma}{0.2}\right)^{3/2} \left(\frac{g(T_f)}{106.75}\right)^{-1/4}\left(\frac{M_{PBH}}{10^{-18} \textsl{g}}\right)^{-1/2},
\end{equation}
where $\mathrm{\beta}$ is the mass fraction of Universe to collapse to PBHs,
$\mathrm{\gamma}$ is the correction factor which depends on the gravitational collapse, $\mathrm{M_{PBH}}$  is the mass of PBHs and $g(T_f)$ is the effective number of degrees of freedom when the PBHs are produced.  The mass of PBHs, $\mathrm{M_{PBH}}$, is associated with the mass inside the Hubble horizon as:
\begin{equation}
M_{PBH}= \gamma M_H=\gamma \frac{4}{3} \pi \rho H^{-3}
\label{43a1}
\end{equation}  
where $\mathrm{\rho}$ is the energy density of the Universe during the collapse. The  $\mathrm{M_{PBH}}$ is given as
\begin{equation}
\label{43}
M_{PBH}=10^{18}  \left(\frac{\gamma}{0.2}\right)  \left(\frac{g(T_f)}{106.75}\right)^{-1/6} \left(\frac{k}{7 \times 10^{13} Mpc^{-1}  }\right)^{-2} \textsl{g}.
\end{equation}
We choose $\mathrm{\gamma=0.8}$ and $\mathrm{g_*=106.75}$ \cite{1975ApJ...201....1C}.

The variance of curvature perturbation $\sigma_\delta$ is related to the power spectrum by the following expression:
\begin{equation}
\label{40}
\sigma_\delta^2 (M_{PBH}(k))= \frac{4(1+w)^2}{(5+3w)^2}  \int \frac{dk' }{k'} \left(\frac{k'}{k}\right)^4  P_R(k') \tilde W^2\left(\frac{k'}{k}\right)
\end{equation}
and for the i-th spectral momentum the smoothed density is give as: 
\begin{equation}
\label{40}
\sigma_i^2 (M_{PBH}(k))= \frac{4(1+w)^2}{(5+3w)^2}  \int \frac{dk' }{k'} \left(\frac{k'}{k}\right)^4 {(k')}^{2i} P_R(k') \tilde W^2\left(\frac{k'}{k}\right)
\end{equation}
\noindent
where   the equation of state,  $w$,  in radiation dominated epoch is equal to $1/3$. $\tilde W \left(\frac{k'}{k}\right)$ is the Fourier transform of the window function.  In the following we assume a Gaussian window function.

The fraction $\mathrm{\beta}$  can be evaluated with the Press Schecter approach (PS) \cite{Press74}  or with the Peak Theory (PT) \cite{Bardeen:1985tr,Young:2014ana,Yoo:2018kvb,Germani:2018jgr}. 
In PS approach the mass fraction, $\mathrm{\beta_{PS}}$, is given by the probability that the overdensity $\mathrm{\delta}$ is above a certain threshold of collapse, denoted as $\mathrm{\delta_c}$. 
To estimate the probability of PBHs formation and establish a connection between the collapse threshold and the power spectrum, we make the assumption that curvature perturbations can be described by Gaussian statistics. This assumption allows us to analyze the formation probability of PBHs and investigate how it relates to the characteristics of the power spectrum.
The fraction $\mathrm{\beta_{PS}}$ for this approach reads as:
\begin{equation}
\label{42}
\beta_{PS}(M_{PBH})= \frac{1}{\sqrt{2 \pi \sigma_\delta ^2 (M)}} \int^{\infty}_{\delta_c} d\delta ~ e^{\frac{-\delta ^2}{2 \sigma_\delta^2(M)}} =\frac{1}{2} \text{Erfc}\left(\frac{\delta_c}{\sqrt{2 }\sigma_\delta}\right).
\end{equation}
\noindent
\noindent
The Eq.~(\ref{42}) is computed using the incomplete gamma function:
\begin{equation}
\beta_{PS}(M_{PBH})=\frac{\Gamma\left(\frac{1}{2}, \frac{\delta_c^2}{2 \sigma_\delta^2}\right)}{2\sqrt{\pi}}. 
\label{42b}
\end{equation}

The PS approach has been found to underestimate the fractional abundances of PBHs by approximately two orders of magnitude \cite{Wang:2021kbh,Stamou:2021qdk}. For this reason we consider the PT as well.   
The peak number density is given by
\begin{equation}
n_{peak}=\int_{\nu_c}^{\infty}\mathcal{N}(\nu)d\nu=\frac{{\sigma_2}^3}{(2\pi)^2(\sqrt{3}\sigma_1)^3}\int^{\infty}_{\nu_c}d\nu \tilde G(\kappa,\nu)e^{-\nu^2/2}
\label{npeak}
\end{equation}
where $\mathrm{\nu} \equiv	\delta /\sigma_0$ and $\mathrm{\nu_c}$ corresponds to  this value at the threshold $\mathrm{\delta_c}$.   The function $\mathrm{\tilde G}$ is given from:
\begin{equation}
\tilde G(\kappa,\nu)=\int_{0}^{\infty}\frac{f(x)}{\sqrt{2\pi(1-\kappa^2)}}\exp\left[ -\frac{(x- \kappa \nu)^2}{2(1-\kappa^2)} \right]dx
\end{equation}
where
\begin{equation}
f=\frac{x^3-3x}{2}\left[ \erf\left(\sqrt{\frac{5}{2}}x\right)+\erf\left(\sqrt{\frac{5}{8}}x\right) \right]+\sqrt{\frac{2}{5\pi}}\left[ \left( \frac{31x^2}{4}+ \frac{8}{5}\right)e^{-5x^2/8}+\left( \frac{x^2}{2}-\frac{8}{5}\right)e^{-5x^2/2} \right]. 
\end{equation}
The mass fraction in PT is given from:
\begin{equation}
\beta_{PT}=\frac{1}{\sqrt{2\pi}}\left(  \frac{{\sigma_2}}{(aH)(\sqrt{3}\sigma_1)} \right)^3\int_{\nu_c}^\infty \tilde G(\kappa,\nu) \exp[-\frac{\nu^2}{2}]d\nu.
\label{eq:beta_integral_pt}
\end{equation}
A recent approximation  gives good results in comparison with the exact numerical PT \cite{Wang:2021kbh}.  According to this approximation the $\mathrm{\beta_{PT}}$ function can be approximated by:
\begin{equation}
\beta_{appr}=\frac{1}{\sqrt{2\pi}}Q^3({\nu_c}^2-1)\exp\left[-\frac{\nu_c^2}{2}\right]
\end{equation}
where $\mathrm{Q=\sigma_1/(aH\sqrt{3}\sigma_\delta)}$.  In the following we will use this approximation. 

As we can notice in this section from the mass fraction $\mathrm{\beta}$   is sensitive to the exact value of the threshold. In both studies of PS and PT,  Eqs.(\ref{42b}) and (\ref{eq:beta_integral_pt}), the value of   $\mathrm{\beta}$ and, hence, the value of the $\mathrm{f_{PBH}}$ is exponentially sensitive to the value of the threshold $\mathrm{\delta_c}$. A slightly different threshold can give different results and that is the reason for the numerical evaluation of the threshold. 



\section{Threshold in inflationary models}

In this section we present the calculation of the threshold $\mathrm{\delta_c}$ for different classes of models proposed in the literature, which are a two-field model with a non-canonical kinetic term, a hybrid model with a waterfal trajectory and a model with an inflection point in the effective scalar potential. 
\textcolor{black}{Specifically, in the hybrid model  the parameters can be combined to  just one parameter, which will be introduced later,  defined as $\Pi$, as it is suggested in \cite{Clesse:2015wea}. Hence, with just one parameter we can derive the height and width of power spectrum.  For the other two models   the parameters which are needed for fixing the width and height of the power spectrum is $\mathrm{b_{1,2}}$ \cite{Braglia:2020eai,Stamou:2021qdk}. The other parameters fix the position of the power spectrum and they are important in order to have a good explanation of the abundances of PBHs to dark matter at the window of $[\mathrm{10^{-16}-10^{-10}}]$ $M\odot$  and simultaneously fitted with the observable constraints of inflation at  CMB scales \cite{Stamou:2021qdk}. However, they do not  affect the height and width.  }
So,  we examine if these models can predict a significant fraction of DM.  We define the parameter of each mechanism, which is responsible for the height and width of power spectrum. Finally, we evaluate the threshold and we propose a value in order these models can explain the maximum of PBHs abundances.

\subsection{Threshold in two field model with non-canonical kinetic term}
In this section we evaluate the threshold and the abundance of PBHs to DM by the  two-field model studied in Ref.~\cite{Braglia:2020eai}. This two-field model is characterized by two stages of inflation with a large non-canonical kinetic coupling, which connects the two fields\cite{Braglia:2020eai,Meng:2022low,Braglia:2020fms,Geller:2022nkr,Braglia:2020taf, Kallosh:2022vha}.  The perturbations  at small scales are dramatically enhanced by the sharp feature in the form of non-minimal coupling.

In particular, the action of a two-field toy model is given by the expression:
\begin{equation}
S=\int d^4 x \sqrt{-g} \left[ \frac{M_P^2}{2}R -\frac{1}{2} (\partial\phi)^2 - e^{-b_1\phi}  (\partial\chi)^2 +V(\phi,\chi) \right]
\label{eq:two_field_action}
\end{equation}
where $\mathrm{M_P}$ is the reduced Planck mass. 
The potential is given as follows:
\begin{equation}
V(\phi,\chi)= V_0 \frac{\phi^2}{\phi_0^2+\phi^2} + \frac{m_{\chi}^2}{2}\chi^2
\end{equation}
where   $\mathrm{\phi}$ is a canonical scalar field, $\mathrm{\chi}$ is a non-canonical scalar field and $\mathrm{b_1}$ an interaction between the
fields.  $\mathrm{V_0}$, $\mathrm{\phi_0}$  and $\mathrm{m_{\chi}}$ are parameters. In the following we consider the  choices of parameters in  \cite{Braglia:2020eai}: 
$\mathrm{V_0/(m_\chi M_P)^2=500}$ and $\mathrm{\phi_0=\sqrt6M_P}$. 


As we mentioned before, for the computation of $\mathrm{\delta_c}$ we need as a first step to evaluate the primordial power spectrum . The power spectrum is evaluated numerically by solving the perturbation of the fields. For this reason we briefly refer to this evaluation. The equations below are written for $\mathrm{n-}$fields, but one can easily obtain those  for one or two fields. More details can be found in \cite{Ringeval:2007am}.

 Generally, the equation of the fields $\mathrm{\varphi^i}$ using efold time is given from (we work in $\mathrm{M_P=1}$):
\begin{equation}
\ddot{\varphi}^c+\Upsilon_{ab}^c\dot{\varphi}^a \dot{\varphi}^b+\left( 3- \frac{1}{2} \dot{\sigma^2}\right)\dot{\varphi}^c+\left( 3- \frac{1}{2} \dot{\sigma^2}\right)\frac{V^c}{V}=0
\end{equation}
where dots represent the derivative in respect to efold time and indices $\mathrm{i=a,b,c}$ represent derivatives with respect to the field.  $\Upsilon^{c}_{ab}$ denotes the Christofel symbol in respect to the fields:
\begin{equation}
\Upsilon^{c}_{ab}=\frac{1}{2}G^{cd}(G_{da,b} +G_{db,a} -G_{ab,d} ),
\end{equation}
where $\mathrm{G_{ij}}$ is the field metric. 
By $\dot\sigma$ we denote the velocity field which is given as follows:
\begin{equation}
\dot{\sigma}^2=G_{ab}\varphi^a\varphi^b.
\end{equation}
The Hubble parameter is given by:
\begin{equation}
H^2=\frac{V}{3-\frac{1}{2}\dot{\sigma^2}}
\end{equation}
and the slow-roll parameter $\epsilon_1$ is given from:
\begin{equation}
{\epsilon}_1=\frac{1}{2}\dot{\sigma^2}.
\label{eq:epsilon_slowroll}
\end{equation}
The equation for the perturbations $\mathrm{\delta\varphi^2}$ of the fields are given by:
\begin{equation}
\begin{split}
&\delta \ddot{\varphi}^c+(3-\epsilon_1)\delta\dot{\varphi}^c+2\Upsilon^c_{ab}\dot{\varphi}^a\delta\dot{\varphi}^b+\left( \Upsilon^{c}_{ab,d}\dot{\varphi}^a\dot{\varphi}^b+\frac{V^{c}_{d}}{H^2} -G^{ca}G_{ab,d}\frac{V^b}{H^2}\right)\delta\varphi^d\\&+\frac{k^2}{a^2H^2}\delta\varphi^c=4\dot\Psi\dot{\varphi}-2\Psi\frac{V^c}{H^2}
\end{split}
\label{eq:perturbation_of _field}
\end{equation}
and the equation for the Bardeen potential $\Psi$  is given by:
\begin{equation}
\ddot{\Psi}+(7-\epsilon_1)\dot{\Psi}+\left( 2\frac{V}{H^2}+\frac{k^2}{a^2H^2}\right)\Psi=-\frac{V_c}{H^2}\delta\varphi^c.
\end{equation}
In this analysis we suppose that we are initially in Bunch-Davies vacuum. 
The power spectrum is given from the equation:
\begin{equation}
P_R=\frac{k^3}{(2 \pi)^2}({R_i}^2)
\label{eq:pr_general}
\end{equation}
where
\begin{equation}
R_i=\Psi +H \sum_i^{N_{fields} } \frac{\dot{\varphi_i} \delta \varphi_i }{\dot{\varphi_i}^2}.
\end{equation}

\begin{figure}[h!]
\centering
\includegraphics[width=80mm,height=61mm]{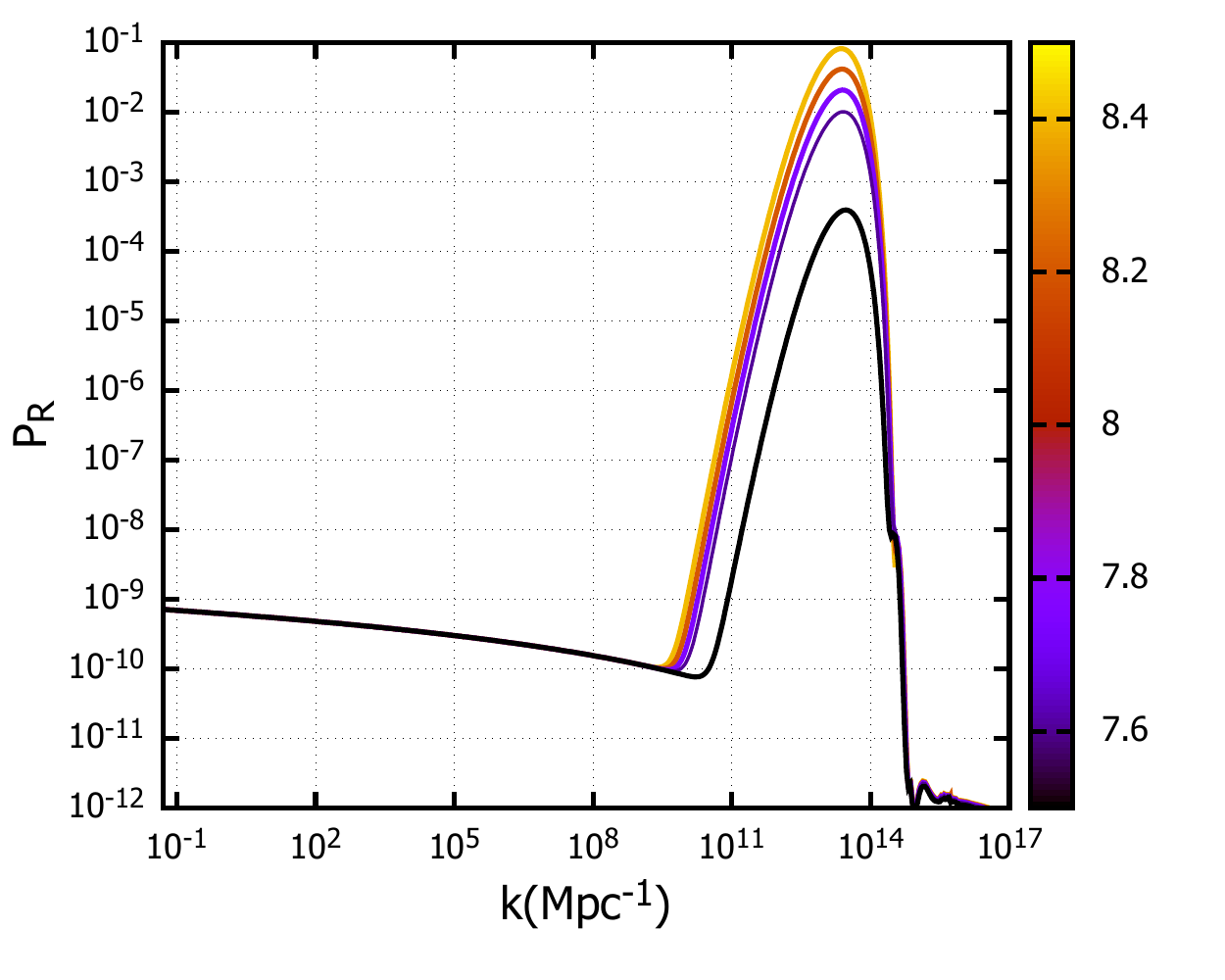}\includegraphics[width=76mm]{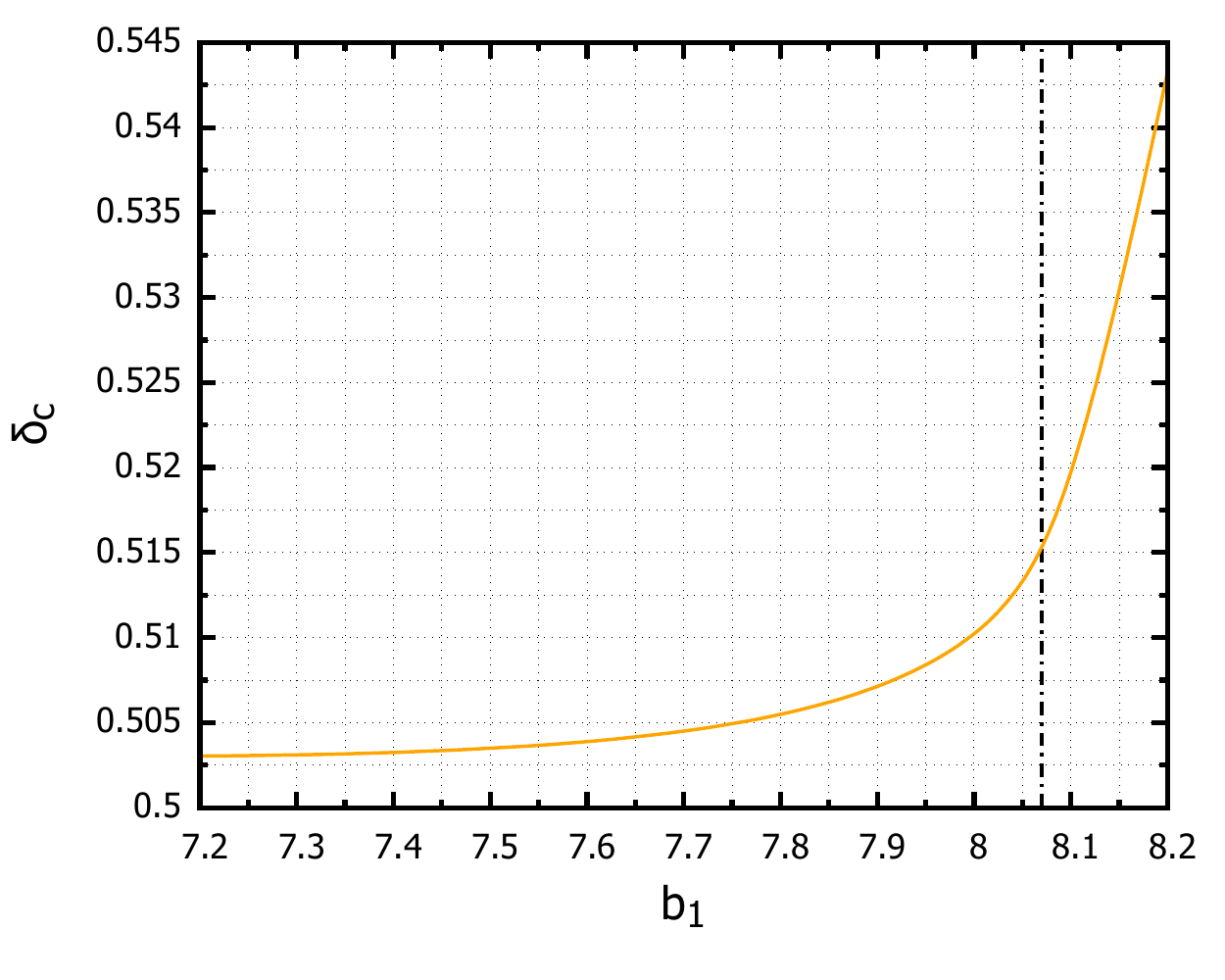}
\caption{ Left: Primordial power spectrum for different values of b. Right: The value of $\delta_{c}$ versus the value of the $\mathrm{b_1}$. The dashed line corresponds the choice of parameter $\mathrm{b_1}$ in order to define all the dark matter in the Universe.}
\label{fig1}
\end{figure}

 In Fig.~\ref{fig1} (left panel) we show the power spectrum for different values of the parameter $\mathrm{b_1}$ given in the kinetic term of the action (\ref{eq:two_field_action}). As one can notice the peak height and the width of the power spectrum depend on the parameter $ \mathrm{b_1}$. For the given power spectra we evaluate the value of the threshold following the steps of the Section \ref{The evaluation of the threshold}. The threshold, which is finally given from the Eq.~(\ref{eq:dc}), is depicted in the right panel of Fig. \ref{fig1} as a function of $ \mathrm{b_1}$ (black line). Therefore one can notice that diferent profiles of power spectrum lead to different thresholds $\delta_c$.

 Finally,  we evaluate the abundance of PBHs to DM using the PT  formalism in Eq. (\ref{eq:abundance}). In the right panel of Fig.~\ref{fig1} we present the specific choice of the value of the $\mathrm{b_1}$ in order to explain the whole amount of DM (orange dashed line).  We observe that this model requires a threshold at around $\mathrm{\delta_c=0.515}$. 
 It is also imperative to verify that the fractional abundance $\mathrm{\Omega_ {PBH}/\Omega_ {DM}}$, given in Eq. (\ref{eq:abundance1}), does not exceed 1. 
In the specific case we have investigated, the value slightly  surpasses 1, leading to the need to explore smaller but notable abundances. Therefore, for obtaining the maximum DM we need a slighter smaller power spectrum and thus a slighter smaller threshold. 
 To sum up, this inflationary mechanism with two fields and a large coupling, which connects these two fields,  can explain a significant amount of  DM from PBHs formation.

\subsection{Threshold in hybrid model}
  Hybrid models for explaining the PBHs have already been proposed \cite{Spanos:2021hpk,Clesse:2015wea}. According to these models, large curvature perturbations can be generated during a mild waterfall trajectory.
In particular, these models are in the framework of two field inflation, where one field  acquires tachyonic solution in the critical point and the other, which plays the role of the inflaton, becomes unstable. The analysis of the background dynamics is decomposed in two phases: a slow-roll phase until the critical point and a second waterfall phase 
till the end of the inflation. The calculation of the inflationary observables during these phases has been performed 
analytically in the slow-roll approximation~\cite{Clesse:2015wea,Kodama:2011vs}.  
   In this section we present the two field hybrid model  \cite{Spanos:2021hpk} for explaining the DM and we evaluate the $\mathrm{\delta_c}$ and then the abundances of PBHs to DM.

The F-term hybrid model in    the globally supersymmetric
renormalizable superpotential reads as follows
 \cite {Copeland:1994vg,Linde:1993cn,Dvali:1994ms}: 
\begin{equation}
W= \kappa \, S  (  \Psi_1 {\Psi}_2  -\frac {M^2}{2})\, ,
\label{spotential}
\end{equation}
where $ \Psi_1, \Psi_2$   are chiral superfields, the scalar component of the  superfield,  $S$ is the gauge singlet  inflaton field, $\kappa$ is a dimensionless  coupling constant and $M$ is a mass.
The scalar potential in hybrid models is given from
\begin{equation}
{  V_F^{\text{SUSY}}= \Lambda\left[ \left(1-\frac{\psi^2}{M^2}\right)^2+ \frac{2 \phi^2 \psi^2}{M^4 }\right]  }\, , 
\label{firstsusy}
\end{equation} 
where we have assumed {$\Lambda=\kappa^2 M^4/4$ . {  In order to fix the non-canonical kinetic term we have $|S|=\phi /\sqrt{2}$ and $|\Psi_1|=| \Psi_2|=\psi/\sqrt{2}$. The potential is flat along the direction $\psi=0$,  $|\phi|>|\phi_c|=M$ and  is given by a constant value for the energy density  $V=\kappa^2 m^4$. The field $\psi$ develops  tachyonic solutions if 
\begin{equation}
\kappa^2(-M^2+\phi^2+6\psi^2)<0 \, . 
\end{equation}
Along the flat direction this  condition  becomes:
\begin{equation}
\phi_c^2< M^2\, .
\label{fcrit}
\end{equation}
The  value of the field $\mathrm{\phi_c}$  denotes the critical point, as below this value the field develops tachyonic solutions.  The inflaton $\mathrm{\phi}$  field moves through the valley until it reaches the critical point. After that the other field, which is  called waterfall, acquires tachyonic solution and the inflaton moves through the waterfall. Finally, the inflation ends in false vacuum.
A potential which leads to enhancement of power spectrum and is derived from hybrid model is given as follows \cite{Clesse:2015wea,Spanos:2021hpk}:
\begin{equation}
{  V_{hybrid}}= \Lambda\left[ \left(1-\frac{\psi^2}{M^2}\right)^2+ \frac{2 \phi^2 \psi^2}{M^4 } +a_1 (\phi- \phi_c)+ a_2 (\phi- \phi_c)^2 \right]
\label{firstsusy}
\end{equation} 
where  $a_\mathrm{1}$ and $a_\mathrm{2}$ are dimensionful parameters. These parameters play a crucial role in shaping the characteristics of the potential and directly influence the behavior of the system at the critical point

The power spectrum of this two-field potential can be analytically estimated by employing the slow-roll approximation and dividing the solution into two distinct phases \cite{Clesse:2015wea,Kodama:2011vs}. 
In the first phase, inflation is solely driven by the inflaton, while the influence of the other field is  considered negligible. However, as the inflationary process progresses, the terms associated with the other field gradually begin to dominate, leading to a significant impact on the dynamics of the system and this signifies the onset of the second phase. The approximated power spectrum  is  given from the following expression \cite{Clesse:2015wea}: 
 \begin{equation}
{ P_R = \frac{1}{4 \pi^2}\frac{\Lambda}{3M_P^2}\left(\frac{  1}{ a_1^2 M_P^4} 
 +\frac{M^4}{64M_P^4 \, \xi_2^2\,  \psi_k^2}\right)    
     \approx \frac{1}{4 \, \pi^2} \frac{\Lambda}{3\, M_P^2} \frac{M^4}{64 \, M_P^4 \, \psi_k^2 \, \xi_2^2}\, .}
     \label{eq:pr_hybrid}
 \end{equation}
 where 
 
 \begin{equation}
\xi_2 = -\frac{ \sqrt{a_1 \chi_2 M}}{2},  \quad\text{with}\quad
 \chi_2 = \ln \Big(   \frac { M^{3/2}\sqrt{a_1}}{2 \psi_0} \Big).
\end{equation} 
 and
\begin{equation}
\psi_k =\psi_0 \, e^{\chi_k},\quad\text{with}\quad \chi_k=\frac{4a_1M_P^4}{M^3}\, (\frac{\chi_2^{1/2} M^{3/2}}{2M_{p}^2\sqrt{a_1}}+ \frac{M  {\phi_c}^{1/2}}{4 \, M_P^2\, a_1^{1/2}\,  x_2^{1/2}}-N)^2.
\end{equation} 
The approximation (\ref{eq:pr_hybrid}) is in good agreement with the numerical evaluation of the power spectrum, as it is shown in Refs.\cite{Spanos:2021hpk,Clesse:2015wea}. For this reason we use this expression for the computation of the thresholds.  Interestingly, it is possible to characterize any of the quantities in this expression by the combination $\mathrm{\Pi}$:
\begin{equation}
\Pi=\frac{M\sqrt{\phi_c}}{M_P^2\sqrt{a_1}}.
\end{equation}
Hence in the next we use this parameter for the evaluation of the threshold. 

\begin{figure}[h!]
\centering
\includegraphics[width=80mm,height=61mm]{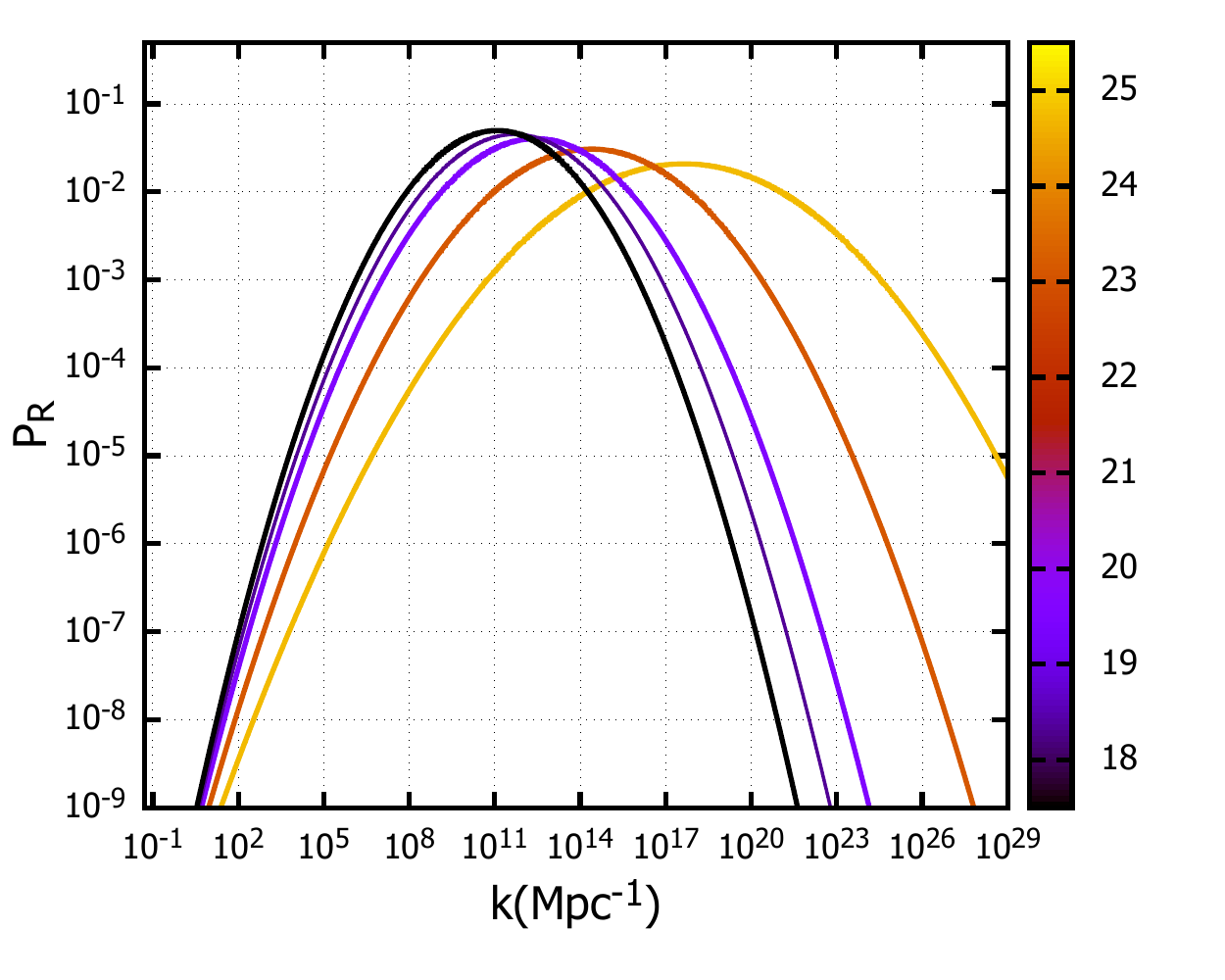}
\includegraphics[width=76mm]{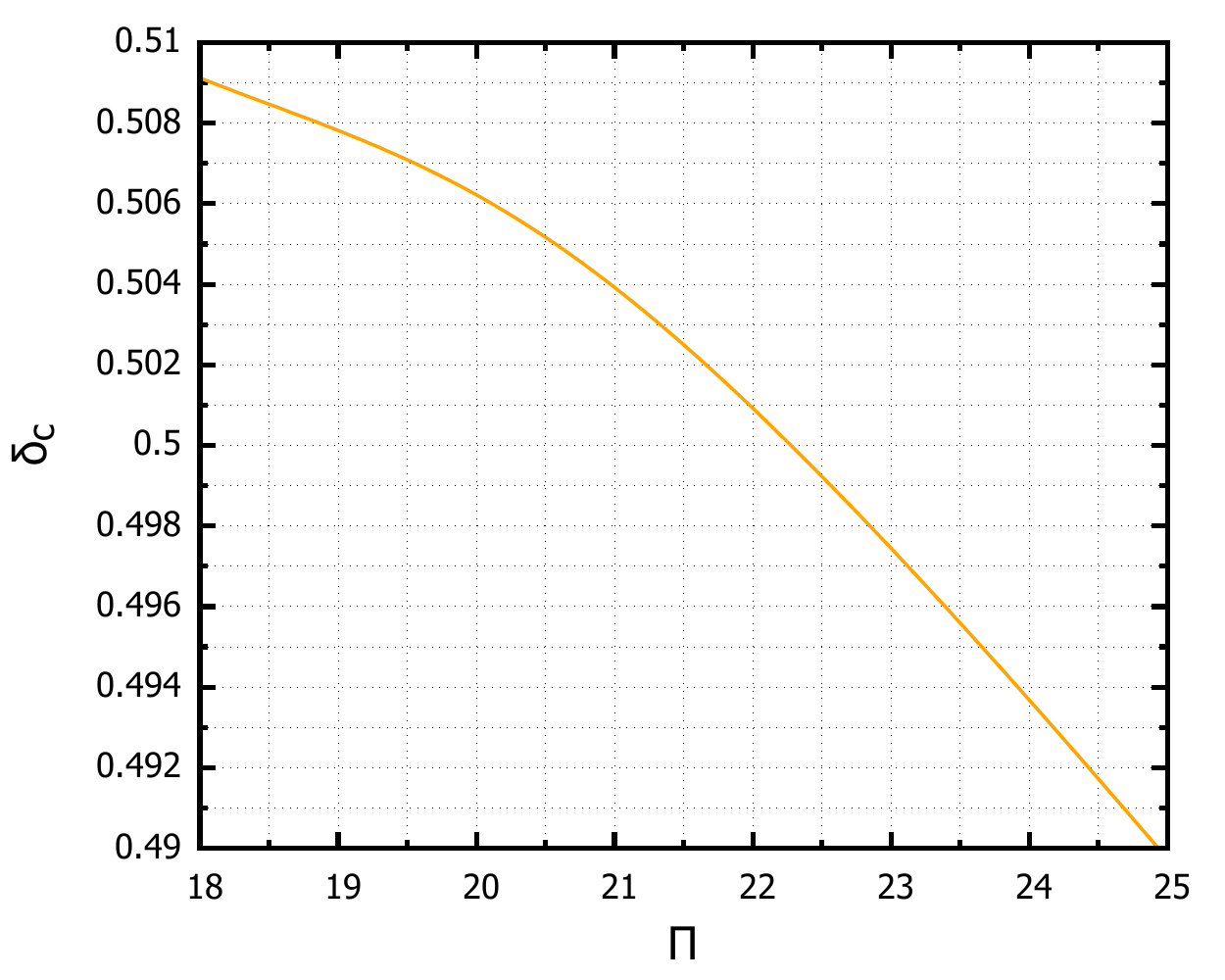}
\caption{ Hybrid model. Left: The power spectrum for different values of $\mathrm{\Pi}$. Right: The value of $\delta_{c}$ versus the value of the $\mathrm{\Pi}$. }
\label{fig2}
\end{figure}

\begin{figure}[h!]
\centering
\includegraphics[width=76mm]{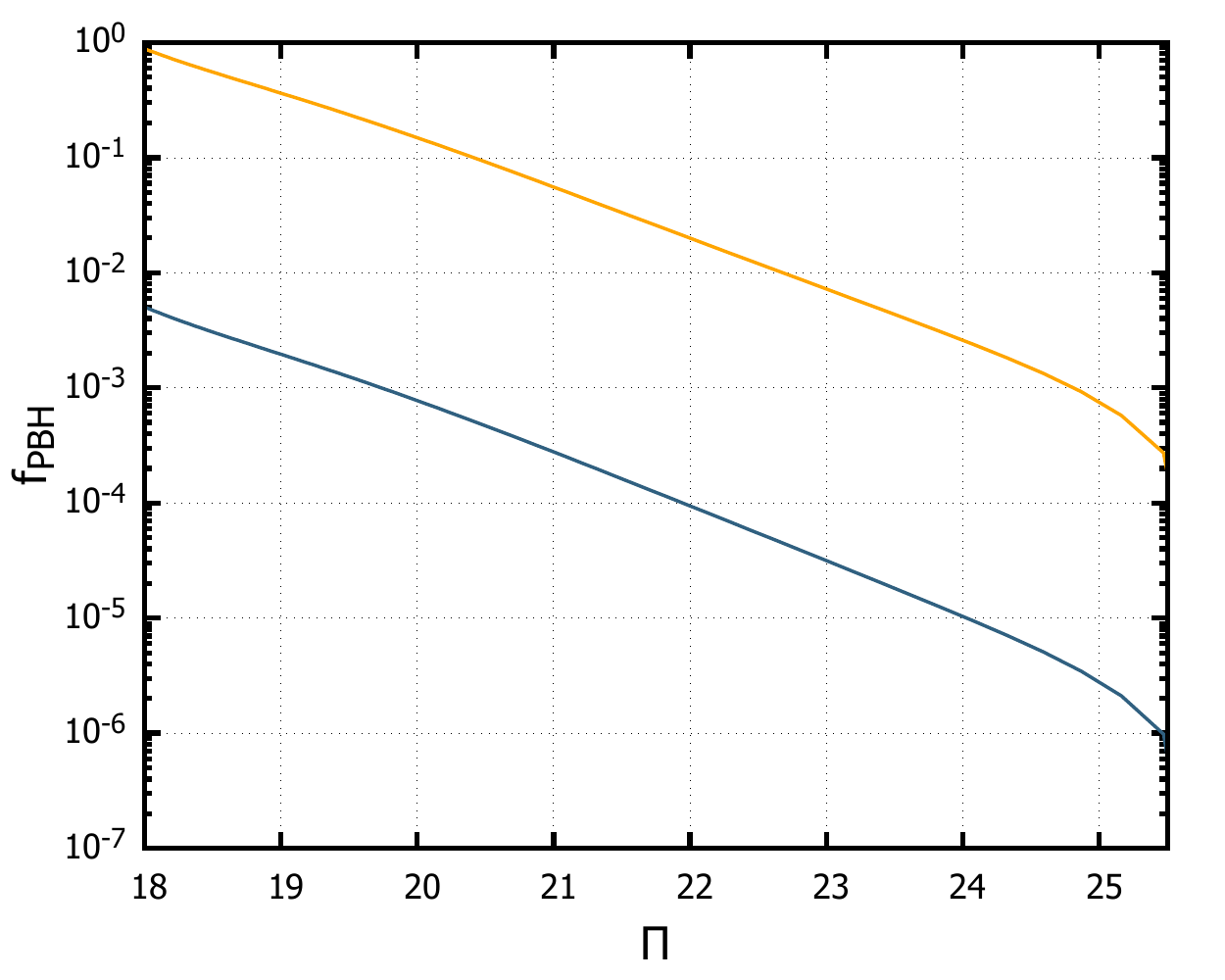}
\caption{  The fraction abundance of PBHs in the case of hybrid models.}
\label{fig3}
\end{figure}

In the left panel of Fig.\ref{fig2} we depict the power spectrum of Eq.(\ref{eq:pr_hybrid}) for different values of the parameter $\mathrm{\Pi}$. We plot the spectra in respect to the comoving wavenumber $\mathrm{k}$, which is connected to the number of efolds as $\mathrm{k=k_*e^{N}}$ and $k_*$ is the pivot scale.  In the right panel of Fig.\ref{fig2} we repeat the calculation for the threshold $\mathrm{\delta_c}$ analyzed in Section \ref{The evaluation of the threshold}.  It is evident that larger values of the power spectrum width lead to a decrease in the corresponding threshold. This result aligns with the findings of Ref. \cite{Musco:2020jjb}, where the study focuses on the Gaussian and lognormal power spectrums. As the width of the power spectrum increases, the shape parameter $\mathrm{q}$ decreases, indicating that the participation of multiple modes in the collapse results in a flatter compactification function~\cite{Musco:2020jjb}.

Through this work we used PT instead of PS approach. To illustrate the contrast between these two methodologies we present a comparison in Fig.\ref{fig3} of how the $\mathrm{f_{PBH}}$ varies with respect to the parameter $\mathrm{\Pi}$. In this figure we display the computed PBH abundance using both the PS (blue line) and PT (orange line) approaches.
Remarkably, a significant disparity emerges between the PS approach and PT. When employing the PS approach, it is shown that we cannot explain a remarkable amount of DM. 
The fraction abundances increase as the value of $\Pi$ decreases. However, in order to have such abundances we need the peak less than $10^{-8}$ and these regions are restricted from  microlensing for Subaru (HSC), Eros/Macho/Ogle \cite{Capela:2013yf,Niikura:2017zjd,Wyrzykowski:2011tr}, ultra faint dwarf (UFD) \cite{Tisserand:2006zx} and CMB measurements \cite{Ali-Haimoud:2016mbv}.  

In contradiction to PS approach,   we observe a notable fraction of $\mathrm{\mathcal{O}(1)}$ with the PT at the values of $\mathrm{\Pi=18}$. These findings highlight the effectiveness of the PT approach, as it offers the opportunity to potentially explain even the whole DM, thus yielding valuable insights for further exploration and analysis. Moreover, it is of utmost importance to ensure that the maximum value of the ratio $\mathrm{\Omega_{PBH}/ \Omega_{DM}}$ does not exceed 1. In this particular instance, we have determined that this value is  lower than one, and so we can elucidate the whole DM content. Therefore, the hybrid models can obtain even the whole DM and the value of the threshold can be at around $\delta_{c}=0.51$.
 

\subsection{Threshold in model with an inflection point}
An intriguing possibility in the framework of inflation is that the existence of an inflection point in the single field potential could provide an explanation for the generation of PBHs and subsequently account for a significant portion of the DM in the Universe. This mechanism arises when the slow-roll parameter, represented by $\mathrm{\epsilon_1}$ in Eq. (\ref{eq:epsilon_slowroll}), acquires a substantial value, resulting in the violation of the slow roll approximation. Importantly, this parameter remains below one, allowing inflation to continue. Following this, a phase emerges where the inflaton remains nearly constant, leading to a local amplification. During this plateau, the power spectrum experiences enhancement, facilitating the production of PBHs during the radiation-dominated phase of the early Universe.
A lot of works have adopted the idea of an  inflection point Refs.\cite{Braglia:2020eai,Clesse:2015wea,Spanos:2021hpk,Ballesteros:2017fsr,Gao:2018pvq,Cicoli:2018asa,Dalianis:2018frf,Garcia-Bellido:2017mdw,Ezquiaga:2017fvi,Nanopoulos:2020nnh,Stamou:2021qdk,Hertzberg:2017dkh,Ballesteros:2019hus,Mahbub:2019uhl}.

In many previous works, a value of $\mathrm{\delta_{c}}$ in the range of acceptance has been used. 	Generally,  one can find the exact value of $\delta_{c}$ for these models and then with fine tuning of the parameters which give the peak height, they can have a significant fraction of DM. For the study of single field inflation with an inflection point we present the model \cite{Stamou:2021qdk}. Similar result one can obtain for other models as well.

The model in Ref.\cite{Stamou:2021qdk} is embedded in no scale supergravity theory. The K\"ahler potential and superpotential are  given from Eqs.:
\begin{equation}
\begin{split}
 K=&-3 \ln\left(1-\frac{|y_1|^2}{3}-\frac{|y_2|^2}{3}\right),\\
 W=&m\left(-y_1y_2 +\frac{y_2y_1^2}{l\sqrt{3}}\right)\left(1+c_2e^{-b_2{y_1}^2}{y_1}^2\right)
 \end{split}
 \label{case2}
\end{equation}
where $\mathrm{y_1}$ and $\mathrm{y_2}$ are chiral fields and $\mathrm{l}$, $\mathrm{b_2}$ and $\mathrm{c_2}$ are parameters. The inflationary direction can be fixed assuming one field  of these fields as the inflaton and the other as the modulo one. The analysis is presented in ~Refs\cite{Ellis:2013nxa,Ellis:2018zya}.
The scalar potential $\mathrm{V}$ can be expressed in terms of the K\"ahler potential, $\mathrm{K}$ and the superpotential $W$:
\begin{equation}
V= e^{K/{M_P}^2}\left[\left( {K^{-1}}\right)^i_{\bar{j}} \left(W^{\bar{j}}+\frac{WK^{\bar{j}}}{{M_P}^2}\right) \left(\bar{W}_i+\frac{\bar{W}K_{i}}{{M_P}^2}\right)- \frac{3|W|^2}{{M_P}^2} \right]
\label{3}
\end{equation}
\noindent
where $( {K^{-1}})^i_{\bar{j}}$ is the inverse of K\"ahler metric and $K^i=\partial K/ \partial \Phi_i$. After the evaluation of the potential and concerning the non-canonical kinetic term of the Lagrangian, 
 one can  obtain an inflection point in the effective scalar potential.  One should also take into account specific values for the parameters in order to obtain significant peaks at small scales in the effective scalar potential.  In this analysis we adopt the parameters of Ref\cite{Stamou:2021qdk}.

In Fig.\ref{fig4} (left panel) we depict the power spectrum as a function of the comoving wavenumber $\mathrm{k}$ for different choice of the parameter b. One can notice that a lot of fine tuning is required for obtaining a capable peak height of the power spectrum. For the evaluation of power spectrum we solve numerically the perturbation of the field, as they given from the differential equations (\ref{eq:perturbation_of _field}) reduced in the single field case. 
Finally we derive the power spectrum  from Eq.~(\ref{eq:pr_general}). 

\begin{figure}[h!]
\centering
\includegraphics[width=70mm]{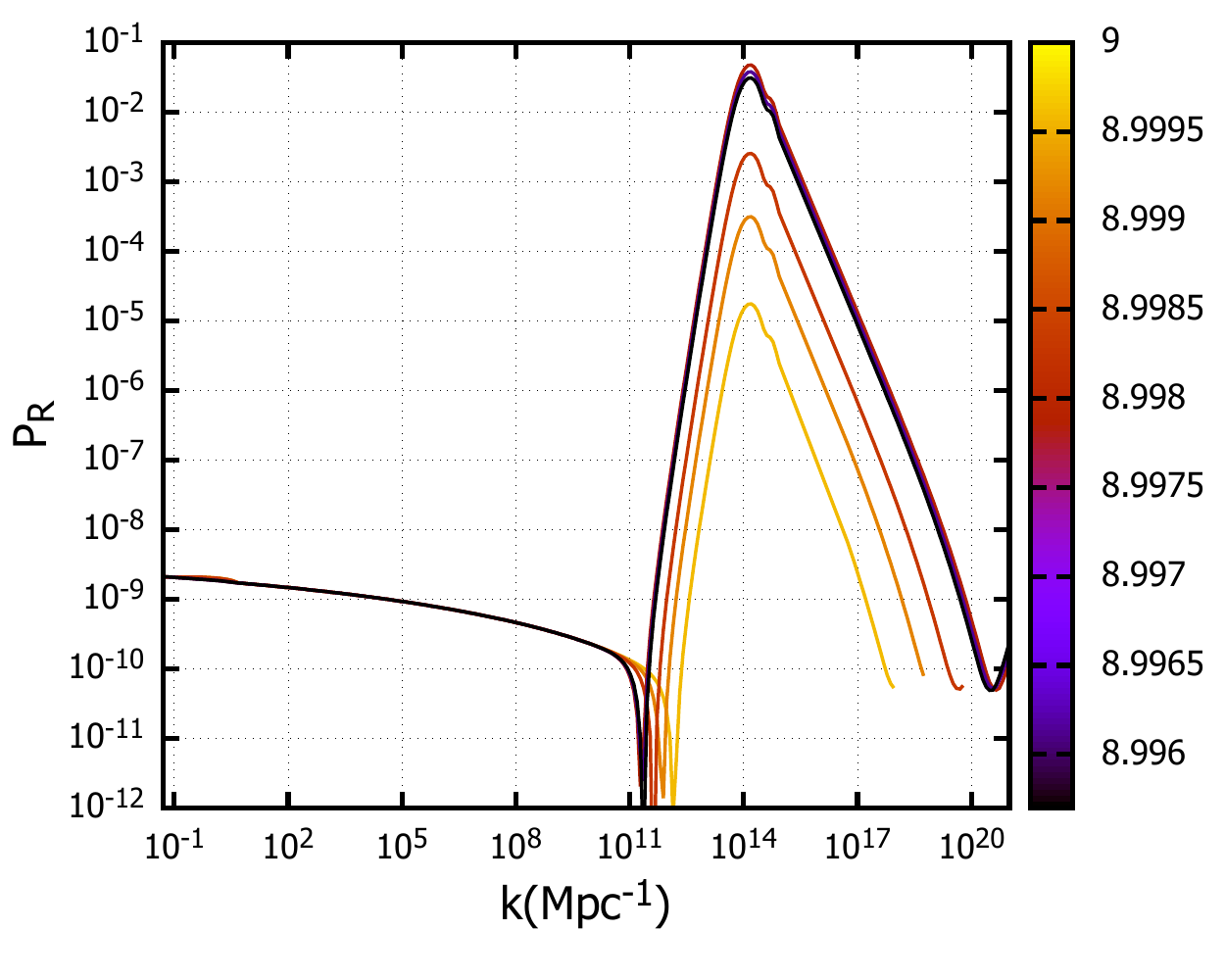}
\includegraphics[width=70mm]{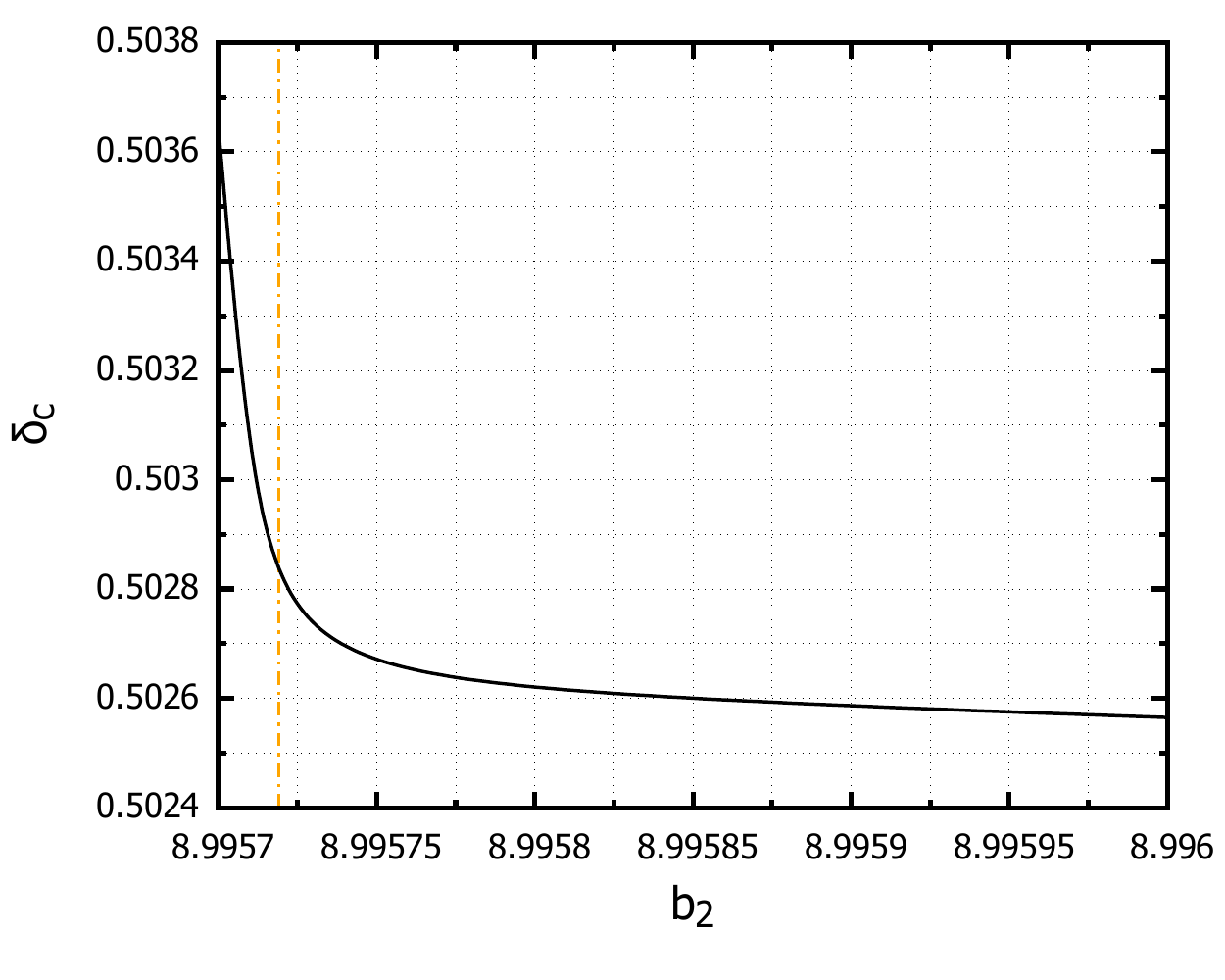}
\caption{ Left: The power spectrum for many choices of parameter $\mathrm{b_2}$. We choose  $l\mathrm{=1.0002}$ and $\mathrm{c_2=14}$. Right: The corresponding thresholds as a function of $\mathrm{b_2}$.}
\label{fig4}
\end{figure}

  In the right panel of Fig.\ref{fig4} we depict the evaluated thresholds $\mathrm{\delta_c}$ for some choices of the parameter $\mathrm{b_2}$ with the formalism from Section \ref{The evaluation of the threshold}. The dashed line in the right panel corresponds to the case where we have $\mathrm{f_{PBH}=1}$. The abundances of PBHs is calculate with the PT, as before. 
  However, it is  important to ensure that the maximum value of the ratio $\mathrm{\Omega_{PBH}/ \Omega_{DM}}$ does not exceed 1.
The results are similar to the case of two field models with a large curvaton field. This value slightly surpasses 1, leading to the need to explore smaller abundances of PBHs.
   Therefore,  inflationary models with an inflection point can  predict a significant amount the DM from PBHs with the value of the threshold at around $\mathrm{\delta_c=0.503   }$. 


Nevertheless, the drawback of these models is that a lot of fine tuning to the interlining parameters is required. An additional problem for these models is that the one loop correction at large scales are too large to accept the validity of perturbation theory \cite{Kristiano:2022maq,Choudhury:2023vuj,Choudhury:2023hvf,Fumagalli:2023hpa}.  

\section{Conclusions}
The formation of PBHs has been the subject of study for decades, with recent advances in numerical simulations providing a more complete understanding of the mechanism involved. The production of PBHs can be explained by a significant enhancement in the scalar power spectrum at small scales. In this work we study the production of PBHs from inflationary models by concerning the evaluation of the threshold $\mathrm{\delta_c}$.  

 Specifically,  we study the evaluation of the threshold $\mathrm{\delta_c}$ in some proposed mechanisms of producing PBHs in the framework of inflation. The first mechanism is based on a two field model with a non-canonical kinetic term, which is responsible for an important enhancement in the scalar power spectrum. The second one is based on a hybrid model characterized by a waterfall trajectory. The last mechanism is a  single inflation with an inflection point in the effective scalar potential. \textcolor{black}{These three models facilitate the computation of the critical threshold by depending on only one parameter, as previously explained. A single parameter dependence is necessary to avoid complex numerical evaluation of thresholds from the power spectrum of a given model.  }
 
 
After evaluating the thresholds for PBHs formation, we  calculate the fractional abundance of PBHs in relation to DM. Two approaches were used for this calculation: the PS approach and the PT. We observed that the PS approach tends to underestimate the results, thus necessitating the adoption of an approximation based on the PT. Our findings indicate that mechanisms based on both two-field inflation and single-field inflation with an inflection point have the ability to account for  a significant fraction  observed DM in the Universe, given specific parameter choices. 

 

Furthermore, we have shown that the mechanism of a mild waterfall in the hybrid model has the capability to account for the entirety of the observed DM content in the Universe. Notably, when comparing the results obtained from the PS approach with those from the PT, it becomes evident that the PS approach yields lower values for the abundances in contrast to the PT approach. Consequently, while the PS approach imposes constraints on these models in order to explain only a fraction of the DM, the PT evaluation of the fraction provides the potential to explain even the entire DM. This discrepancy underscores the enhanced effectiveness and broader explanatory power of the PT approach in understanding the origins of DM in the Universe.

Finally, it is worth noting that in all the investigated models, the threshold value consistently converges around $\mathrm{0.51}$. This consistent threshold value holds significant implications for future comparative studies focused on the production of PBHs through inflationary processes. Understanding the precise threshold value provides a crucial benchmark for assessing the viability and efficiency of different inflationary scenarios in generating PBHs. These findings pave the way for more comprehensive and in-depth investigations into the underlying mechanisms that contribute to the formation of PBHs.



\section*{Acknowledgments}
We would like to express our gratitude to S. Clesse, A. Escriva, and I. Musco for their fruitful discussions and valuable insights. Their contributions have greatly enriched this  work. I.D.S. acknowledges funding by the Belgian Fund for Research F.R.S.-FNRS through an Incentive Grant for Scientific Research (MIS).

\bibliographystyle{elsarticle-num}
\bibliography{bib}

\end{document}